\documentclass{article}
\usepackage{spconf,amsmath,graphicx}

\usepackage{color}
\usepackage{cite}
\usepackage{amssymb,amsfonts,url,amsmath}
\usepackage{algorithmic}
\usepackage{textcomp}
\usepackage[rgb]{xcolor}
\usepackage{float}
\usepackage{placeins}
\newcommand{\stz}{\rule{0mm}{2.3ex}}

\usepackage{bm}
\usepackage{color,soul,cite}
\usepackage{MnSymbol}
\usepackage{wasysym}
\usepackage{multirow}
\usepackage{makecell}
\usepackage{colortbl,hhline}
\usepackage[super]{nth}
\usepackage{subcaption}
\usepackage{adjustbox}
\usepackage{booktabs,pifont}

\newcommand{\xmark}{\ding{55}}
\newcommand{\cmark}{\ding{51}}

\title{Employing Real Training Data for Deep Noise Suppression}
\name{Author(s) Name(s)}
\address{Author Affiliation(s)\\
author@affiliation.tld}

\iftrue\name{Ziyi Xu, Marvin Sach, Jan Pirklbauer, Tim Fingscheidt}
\address{Institute for Communications Technology, Technische Universit{\"a}t Braunschweig, Germany\\
	$\left \{ \text{ziyi.xu, m.sach, jan.pirklbauer, t.fingscheidt} \right \}$@tu-bs.de\\       
}\fi

\begin{document}
\ninept
\maketitle
\begin{abstract}
Most deep noise suppression (DNS) models are trained with ref\-erence-based losses requiring access to clean speech. However, sometimes an additive microphone model is insufficient for real-world applications. Accordingly, ways to use real training data in supervised learning for DNS models promise to reduce a potential training/inference mismatch. Employing real data for DNS training requires either generative approaches or a reference-free loss without access to the corresponding clean speech. In this work, we propose to employ an end-to-end non-intrusive deep neural network (DNN), named \texttt{PESQ-DNN}, to estimate perceptual evaluation of speech quality (PESQ) scores of enhanced real data. It provides a reference-free perceptual loss for employing real data during DNS training, maximizing the PESQ scores. Furthermore, we use an epoch-wise alternating training protocol, updating the DNS model on real data, followed by \texttt{PESQ-DNN} updating on synthetic data. The DNS model trained with the \texttt{PESQ-DNN} employing real data outperforms all reference methods employing only synthetic training data. On synthetic test data, our proposed method excels the Interspeech 2021 DNS Challenge baseline by a significant 0.32 PESQ points. Both on synthetic and real test data, the proposed method beats the baseline by 0.05 DNSMOS points – although \texttt{PESQ-DNN} optimizes for a different perceptual metric.
\end{abstract}
\begin{keywords}
speech enhancement, denoising, non-intrusive PESQ estimation, real recordings
\end{keywords}
\section{Introduction}
\label{sec:intro}

Deep noise suppression (DNS) methods employ deep neural networks (DNNs) to improve perceptual quality and intelligibility of a speech signal distorted by background noise and sometimes reverberation. Most DNS models are trained with reference-based loss functions requiring access to clean speech, in some cases even to additive noise. Accordingly, the employed training data is synthetic: The clean speech and noise are pre-processed separately and mixed by addition to simulate noisy mixtures with different signal-to-noise ratios (SNRs), while the effect of reverberation (if considered) is realized by convolution with either simulated or recorded room impulse responses \cite{pascual2017segan,pandey2018adversarial,wang2018supervised,tan2019complex,strake2019separated,strake2020fully,strake2020DNS,braun2021consolidated}. Strake et al.\ \cite{strake2020fully} proposed a fully convolutional recurrent neural network ({\tt FCRN}) trained with a loss function based on the mean squared error (MSE) for joint denoising and dereverberation. Braun et al.\ \cite{braun2021consolidated} investigated the contributions of various MSE-based spectral losses for training a recurrent neural network (RNN) and proposed a loss function combining complex spectral and compressed magnitude loss terms, which provides state-of-the-art performance. 

Beyond training, synthetic data is also prominently used to evaluate the performance of the trained DNS model. This employs intrusive instrumental metrics requiring access to the clean reference speech signal, e.g., perceptual evaluation of speech quality (PESQ) \cite{ITUT_pesq_wb_corri} and perceptual objective listening quality assessment (POLQA) \cite{ITUT_polqa_2018}, both reflecting the perceived speech quality, and short-time objective intelligibility (STOI) \cite{taal2010short} for estimating the speech intelligibility. However, the linear additive microphone model with statistically independent speech and noise is often invalid in real-world applications due to, e.g., the Lombard effect, where the speakers attend to speak loudly with higher pitch under very noisy conditions, and also due to clipping effects of the overloaded high-level microphone signal. Such effects may lead to a potential training/inference mismatch.

In contrast, evaluation of a DNS model on real test recordings requires either a discriminator from a GAN-based approach \cite{pascual2017segan,pandey2018adversarial}, or non-intrusive instrumental metrics without knowing the clean reference signal. In Microsoft DNS Challenges \cite{reddy2021icassp,reddy2021interspeechIN}, Reddy et al.\ provided a non-intrusive instrumental measure called DNSMOS \cite{reddy2021dnsmos} to predict ITU-T P.808 \cite{ITU_P808} subjective rating scores employing a DNN. The latest DNSMOS \cite{reddy2022dnsmos} separately estimates the quality of the speech component, background noise, and the overall enhanced speech following ITU-T P.835 \cite{ITUT_P835}. Mittag et al.\ proposed an RNN named {\tt NISQA} \cite{mittag2021nisqa} to estimate the subjective rating scores focusing on coded speech under various transmission conditions. In \cite{xu2021inter,xu2021deepT,xu2022iwaenc}, an end-to-end non-intrusive {\tt PESQNet} DNN was proposed to estimate the PESQ scores of the enhanced speech signal, thus providing a reference-free perceptual loss for training a DNS model. In \cite{xu2021inter}, the authors proposed a joint loss combining an MSE loss and the perceptual loss offered by the {\tt PESQNet} for fine-tuning a pre-trained DNS model employing both {\it real} and {\it synthetic} data. However, this approach revealed instabilities in the {\tt PESQNet}/DNS joint training. The same authors solved the problem with an epoch-based alternating training protocol in their following works \cite{xu2021deepT,xu2022iwaenc}. Notably, the authors mentioned that the new approach ``opens the door'' for using real training data, however, without providing evidence for this hypothesis. In \cite{xu2023PESQDNN}, an improved end-to-end non-intrusive {\tt PESQ-DNN} was proposed to estimate PESQ scores for coded speech signals obtained from various wideband codecs considering different transmission conditions. {\tt PESQ-DNN} employs a complex spectrogram as input and yields better performance than the {\tt PESQNet} and a more stable training process.

In this work, our contributions are fourfold: First, we successfully entered the door "opened" in \cite{xu2021deepT,xu2022iwaenc} to use real training data with their proposed training protocol. Second, we replace {\tt PESQNet} with the better performing {\tt PESQ-DNN}. Third, we investigate the potential of employing a {\it minibatch}-based {\tt PESQ-DNN}/DNS alternating training protocol. Finally, we train the {\tt FCRN} proposed in \cite{strake2020fully} as our DNS model and perform an extensive experimental evaluation. Even though we optimize for PESQ during {\tt FCRN} training, we provide the results showing that the use of real training data improves {\it both} on synthetic and real test data. It improves {\it both} on PESQ and on other non-intrusive instrumental metrics, e.g., DNSMOS.

The rest of the paper is structured as follows: Section 2 introduces the DNS and the employed {\tt PESQ-DNN} processing. The novel {\tt PESQ-DNN}/DNS joint training employing real training data is presented in Section 3. The experimental setup and results are discussed in Section 4, followed by the conclusion in Section 5.

\section{Speech Enhancement With PESQ-DNN}
\vspace*{-2mm}
\subsection{Deep Noise Suppression (DNS)}
Following \cite{xu2021inter,xu2021deepT,xu2022iwaenc}, we pre-train the {\tt FCRN} proposed in \cite{strake2020fully} as our DNS model with synthetic data. Accordingly, the microphone mixture $y(n)$ is synthesized from the clean speech signal $s(n)$, reverberated by the room impulse response (RIR) $h(n)$, and disturbed by additive noise $d(n)$ as
\vspace*{-1mm}
\begin{equation} \label{micro_mixture}
\vspace*{-1mm}
y(n)=s(n)*h(n)+d(n)= s^\text{rev}(n)+d(n),
\end{equation}
with $s^\text{rev}(n)$ and $n$ being the reverberated clean speech signal and the discrete-time sample index, respectively, and $*$ denoting a convolution operation. Subsequently, after windowing, the signals are converted to the discrete Fourier transform (DFT) domain by:
\begin{equation} \label{micro_fft}
Y_\ell(k)=S^\text{rev}_\ell(k)+D_\ell(k),
\vspace*{-1mm}
\end{equation}
with frame index $\ell$ and frequency bin index $k\!\in\!\mathcal{K}\!=\!\left \{0,1,\ldots,K\!-\!1\right \}$, and $K$ being the DFT size. The employed {\tt FCRN} implicitly estimates a magnitude-bounded complex mask $M_\ell\left(k \right )\in\mathbb{C}$, with $\left|M_\ell\left(k \right )\right|\in\left [ 0,1 \right ]$, to enhance the noisy speech spectrum by:
\begin{equation} \label{clean_speech_est}
\vspace*{-1mm}
\hat{S}_\ell\left (k \right )=Y_\ell(k)\cdot M_\ell\left(k \right ).
\vspace*{-1mm}
\end{equation}
Finally, the enhanced speech spectrum $\hat{S}_\ell\left (k \right )$ is subject to an inverse DFT (IDFT), followed by overlap add (OLA) to reconstruct the time-domain enhanced speech signal $\hat{s}(n)$.

Instead of the joint denoising and dereverberation loss proposed in \cite{strake2020DNS} and used in \cite{xu2021inter,xu2021deepT,xu2022iwaenc}, we employ the loss proposed by Braun et al.\ \cite{braun2021consolidated} for pre-training, proven to offer state-of-the-art performance:
\begin{equation} \label{Braun}
\begin{aligned}
J^\text{Braun}_u\!&=\!\frac{\alpha}{L_u\!\cdot\!K}\!\sum_{\ell\in\mathcal{L}_u}\sum_{k\in\mathcal{K}}\!\left|\hat{S}_\ell(k)\!-\!S_\ell(k)\right|^2\\
&+\!\frac{1\!-\!\alpha}{L_u\!\cdot\!K}\!\sum_{\ell\in\mathcal{L}_u}\sum_{k\in\mathcal{K}}\!\left|\left|\hat{S}_\ell(k)\right|^c\!-\!\left|S_\ell(k)\right|^c\right|^2,
\end{aligned}
\end{equation}
with $\ell\!\in\!\mathcal{L}_u\!=\!\left \{1,2,\ldots,L_u\right \}$, and $L_u$ being the number of frames in an entire utterance indexed by $u$. Following \cite{braun2021consolidated}, the compression and the weighting factors between the complex and magnitude loss terms are set to $c=0.3$ and  $\alpha=0.7$, respectively.

\vspace*{-2mm}
\subsection{Non-Intrusive \texttt{PESQ-DNN}}
\vspace*{-2mm}
To estimate PESQ scores of the enhanced speech signal obtained from our {\tt FCRN} DNS, we adopt the non-intrusive {\tt PESQ-DNN} proposed by Xu et al.\ \cite{xu2023PESQDNN}, which employs frame-level embeddings (FLE) and average pooling. The input of the non-intrusive {\tt PESQ-DNN} is the enhanced speech spectrum $\hat{S}_\ell\left (k \right )$, with $\ell\!\in\!\mathcal{L}_u$ defined in \eqref{Braun}. The final output is the utterance-wise estimated PESQ score $\widehat{\text{PESQ}}_u$, which should be close to its corresponding ground truth $\text{PESQ}_u$ measured by ITU-T P.862.2 \cite{ITUT_pesq_wb_corri}. For training the {\tt PESQ-DNN}, we employ the loss function \cite{xu2023PESQDNN}
\begin{equation} \label{PESQ_FLE}
J^\text{PESQ}_u\!=\left(\widehat{\text{PESQ}}_u\!-\!\text{PESQ}_u\right)^2\!+\tfrac{\alpha_u}{B_u\cdot L_b}\!\sum\limits_{b\in\mathcal{B}_u}\sum\limits_{\ell\in\mathcal{L}_b}\!\left(q_{b}(\ell)\!-\!\text{PESQ}_u\right)^2,
\end{equation}
with $q_{b}(\ell)$ being some intermediate FLE PESQ scores, representing the predicted PESQ scores for each frame indexed by $\ell\!\in\!\mathcal{L}_b$, and $\mathcal{L}_b$ being the set of frames belonging to feature block $b$. Parameter $B_u=\left|\mathcal{B}_u\right|$ represents the total number of feature blocks for utterance $u$. The utterance-wise weighting factor is represented by 
\begin{equation} \label{weights}
\alpha_u= 0.9^{\left|\text{PESQ}_u-\text{PESQ}_\text{max}\right|},
\end{equation}
with $\text{PESQ}_\text{max}=4.64$ being the maximum PESQ score defined in \cite{ITUT_pesq_wb_corri}. This weighting factor encourages the intermediate FLE PESQ to be equal to the utterance-wise PESQ score for a speech utterance with good quality: A perfect overall perceptual quality should be reflected everywhere in the utterance, e.g., each frame should have the same high PESQ score.

\vspace*{-2mm}
\section{DNS Training with Real Recordings}
\begin{figure}[t!]
\centering
\includegraphics[width=0.44\textwidth]{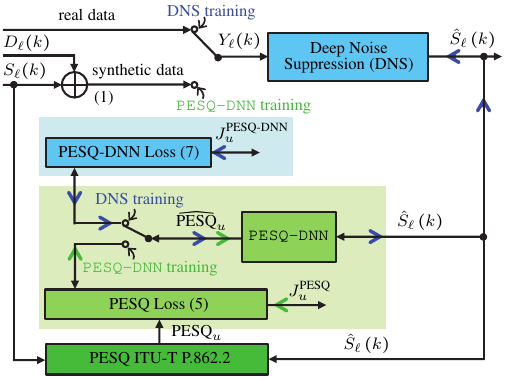}
\caption{{\bf \texttt{PESQ-DNN} and DNS joint training setup.} The DNS and the {\tt PESQ-DNN} are trained alternatingly, controlled by the two switches locked together, both either in the upper or lower position. Colored arrows: gradient flow.}
\label{system}
\end{figure}

\vspace*{-2mm}
Following the training schemes proposed in \cite{xu2021deepT,xu2022iwaenc}, we employ {\tt PESQ-DNN} to fine-tune the pre-trained {\tt FCRN} DNS, however, employing {\it real} data to further increase the PESQ scores of the enhanced speech signal. Accordingly, we define a loss
\begin{equation} \label{real}
\vspace*{-1mm}
J^\text{PESQ-DNN}_u\!=\left(\widehat{\text{PESQ}}_u-\text{PESQ}_\text{max}\right)^2
\vspace*{-1mm}
\end{equation}
for utterance $u$ provided by the {\tt PESQ-DNN} to maximize the PESQ scores of the {\tt FCRN} DNS, with $\text{PESQ}_\text{max}\!=\!4.64$.
We investigate two different joint training protocols, as depicted in Fig.\,\ref{system}. To keep the employed {\tt PESQ-DNN} up-to-date, we adopt the {\it epoch} (EP)-level joint training protocol, originally proposed in \cite{xu2021deepT}: The {\tt FCRN} DNS is fine-tuned with a fixed {\tt PESQ-DNN} for one epoch of {\it real} data, followed by {\tt PESQ-DNN} training employing an epoch of {\it synthetic} data, adapting to the current updated {\tt FCRN} DNS. Please note that for {\tt PESQ-DNN} training we can only apply {\it synthetic} data, since only then the clean reference signal is available for target PESQ score calculation following ITU-T P.862.2 \cite{ITUT_pesq_wb_corri}.

Besides the EP-level alternating training protocol successfully used in \cite{xu2021deepT,xu2022iwaenc}, we also investigate a {\it minibatch} (MB) -level training protocol, where the {\tt FCRN} DNS is fine-tuned with one minibatch of {\it real} data followed by the {\tt PESQ-DNN} updating with one minibatch of {\it synthetic} data. The DNS and {\tt PESQ-DNN} are trained alternatingly on {\it real} and {\it synthetic} data, controlled by both switches in Fig.\,\ref{system} in the upper and lower positions, respectively. The blue arrows indicate the gradient flow back-propagated for the DNS training, while the green ones are for {\tt PESQ-DNN}.

\section{Experimental Evaluation}
\subsection{Datasets, Methods, Training, and Metrics}
\begin{table*}[!t]
\centering
\caption{Results on the \textbf{synthetic development set} $\mathcal{D}^\mathrm{dev}_\mathrm{DNS1}$. DNS fine-tuning data may be synthetic (``syn") and/or real (``real"). DNSMOS refer to the overall (OVRL) metric. Our new method is shown for the minibatch (MB) or epoch (EP) alternating training protocol. Best results are in {\bf bold} font, and the second best are \underline{underlined}. Proposed method with $*$.}
\ninept
\setlength\tabcolsep{3pt}
\vspace*{-2mm}
\begin{tabular}{c c c c c c c c c c c c}
	\hline
	&\multirow{2}{*}{Methods} & \multicolumn{2}{c}{\stz{DNS fine-tuning}}& \multicolumn{4}{c}{\stz{Without reverb}}& \multicolumn{4}{c}{\stz{With reverb}}\\ \cmidrule(l{.25em}r{.25em}){3-4} \cmidrule(l{.25em}r{.25em}){5-8} \cmidrule(l{.25em}r{.25em}){9-12} 
	&   & \multicolumn{1}{c}{syn} & \multicolumn{1}{c}{real} & PESQ & DNSMOS &  \stz{STOI}  &   $\Delta\text{SNR}_\text{seg}$[dB]   & PESQ & DNSMOS & \stz{STOI} &  SRMR \\ \hline
	\multirow{2}{*}{}& \multicolumn{1}{l}{\stz{Clean}}  &    -  &  -   &  4.64 &  3.27 &  1.00   &  -  &  4.64  & 3.27  & 1.00   &-  \\ \hhline{~~~~~~~~~~~~}
	& \multicolumn{1}{l}{Noisy}  &    -  &  -   &  \stz{2.21} &  2.47 &  0.91   &  -  &  1.57  & 1.42  &0.56    &-  \\ \hline
	\multirow{7}{*}{\rotatebox{90}{REF \quad}}& \multicolumn{1}{l}{\stz{DNS3 Baseline \cite{braun2020data}}}      &    \cmark  &  \xmark         &     3.15   &  3.08    &    0.94   &    {6.30} & {1.68}   &  2.28   & \underline{0.62}   & 6.33\\ \hhline{~~~~~~~~~~~~}
	& \multicolumn{1}{l}{\stz{{\tt FCRN} \cite{strake2020DNS}}} &    \cmark  &  \xmark  &      {3.37}   &  3.09   &    {\bf 0.96}   &    8.35 & {\bf 1.95} &  2.17    &  {\bf 0.63}  & {7.25}\\ \hhline{~~~~~~~~~~~~}
	&\multicolumn{1}{l}{\stz{{\tt FCRN}/{\tt PESQNet} \cite{xu2021inter}}}  &    \cmark  &  \cmark  &    3.29  &  \underline{3.18}  & \underline{0.95}  &    {8.52}  & {1.92}  &  \underline{2.29}   &  \underline{0.62}  & 6.85 \\ \hhline{~~~~~~~~~~~~}
	&\multicolumn{1}{l}{\stz{{\tt FCRN}/{\tt PESQNet}, EP \cite{xu2021deepT}}} & \cmark  &  \xmark   &    {3.45}  &  3.12  & {\bf 0.96}  &    {8.48}  & {\bf 1.95}  &  2.19   &  \underline{0.62}  & {7.38} \\ \hhline{~~~~~~~~~~~~}
	&\multicolumn{1}{l}{\stz{{\tt FCRN}/{\tt MF-PESQNet}, EP \cite{xu2022iwaenc}}} &  \cmark  &  \xmark  &    3.47  &  3.12   &    {\bf 0.96}  & 8.54  & {\bf 1.95}  &  2.18   &  \underline{0.62}  & {\bf 7.53} \\ \hhline{~~~~~~~~~~~~}
	&\multicolumn{1}{l}{\stz{{\tt FCRN}, $J^\text{Braun}_u$}} & \cmark  &  \xmark  &    \underline{3.58}  &  {3.17}   &    {\bf 0.96}  & \underline{9.10}  & \underline{1.94}  &  2.27   &  \underline{0.62}  & \underline{7.41} \\ 
	&\multicolumn{1}{l}{\stz{{\tt FCRN}, $J^\text{Braun}_u$, cont.}} & \cmark  &  \xmark  &    3.57  &  3.17   &   {\bf 0.96}  & 9.00  & 1.90  &  2.26   &  \underline{0.62}  & 7.30 \\ \hline
	\multirow{2}{*}{\rotatebox{90}{NEW}} &\multicolumn{1}{l}{\stz{{\tt FCRN}/{\tt PESQ-DNN}, MB}}&    \xmark  &  \cmark   &    3.30  &  3.15  & \underline{0.95}  &    7.39  & 1.79  &  2.22   &  0.59  & 6.14 \\ \hhline{~~~~~~~~~~~~}
	&\multicolumn{1}{l}{\stz{{\tt FCRN}/{\tt PESQ-DNN}, EP$^*$}}&    \xmark  &  \cmark   &    {\bf 3.60}  &  {\bf 3.20}   &    {\bf 0.96}  & {\bf 9.11}  & \underline{1.94}  &  {\bf 2.31}   &  \underline{0.62}  & 7.38 \\ \hline
	
\end{tabular}
\label{DNS1_dev}
\vspace*{-2mm}
\end{table*}

In this work, signals have a sampling rate of $16\,\text{kHz}$ and we apply a periodic Hann window with a frame length of $384$ with a $50\%$ overlap, followed by an FFT of size $K=512$. The employed {\tt FCRN} DNS is adopted from \cite{strake2020fully} and is used in \cite{strake2020DNS, xu2021inter,xu2021deepT,xu2022iwaenc}. The number of input and output frequency bins of the employed {\tt FCRN} DNS is set to $260$. The used {\tt PESQ-DNN} has exactly the same topology as originally proposed in \cite{xu2023PESQDNN}, employing the enhanced speech spectra (containing $260$ frequency bins) from the {\tt FCRN} DNS as input. The last three frequency bins of the inputs for both {\tt FCRN} DNS and {\tt PESQ-DNN} are redundant due to divisibility constraints. For the {\tt PESQ-DNN}, the number of frames in each feature block, denoted by $L_b=\left|\mathcal{L}_b\right|$ in \eqref{PESQ_FLE}, is set to $16$.

The {\tt FCRN} DNS and {\tt PESQ-DNN} are firstly pre-trained and validated on {\it synthetic} datasets $\mathcal{D}^\text{train}_\text{WSJ0}$ and $\mathcal{D}^\text{val}_\text{WSJ0}$, respectively, which are used in \cite{xu2021deepT, xu2022iwaenc} with clean speech from WSJ0 speech corpus \cite{Garofalo2007} and noise from DEMAND \cite{thiemann2013diverse} and QUT \cite{dean2010qut}. The fine-tuning of the {\tt FCRN} DNS and {\tt PESQ-DNN} are based on the Microsoft DNS Challenge dataset, which features harsh conditions considering the diversities of languages, speakers, noise types, and SNR conditions. Accordingly, we follow the two-stage fine-tuning strategy applied in \cite{xu2021deepT, xu2022iwaenc}. The \nth{1}-stage fine-tuning of the {\tt FCRN} DNS and {\tt PESQ-DNN} employing {\it synthetic} dataset based on the official Interspeech 2021 DNS Challenge (dubbed DNS3) training material \cite{reddy2021interspeechIN}, including $100$ hours of training set $\mathcal{D}^\text{train}_\text{DNS3}$ and $10$ hours of validation set $\mathcal{D}^\text{val}_\text{DNS3}$ generated with the same setting as in \cite{xu2021deepT}. 

In pre-training and \nth{1}-stage fine-tuning, the {\tt FCRN} DNS and {\tt PESQ-DNN} are trained using losses \eqref{Braun} and \eqref{PESQ_FLE} with an initial learning rate of $10^{-4}$ and $5\cdot10^{-5}$, respectively, employing Adam optimizer. The learning rate is halved once the corresponding validation loss does not improve for two consecutive epochs. The training is stopped after the validation loss is not improving for five consecutive epochs, and the models with the lowest validation loss are saved. Accordingly, the \nth{1}-stage fine-tuned {\tt FCRN} DNS serves as a reference method, denoted as ``{\tt FCRN}, $J^\text{Braun}_u$" in the following discussion, to illustrate the PESQ performance improvement obtained from the proposed \nth{2}-stage fine-tuning employing {\it real} data. We pre-train and fine-tune the same {\tt FCRN} DNS employing the joint denoising and dereverberation loss proposed in \cite{strake2020DNS} as one of the baselines, shown as ``{\tt FCRN} \cite{strake2020DNS}" in Tabs.\,\ref{DNS1_dev}, \ref{DNS2_test}, and \ref{DNS3_dev}.

The \nth{2}-stage fine-tuning builds upon ``{\tt FCRN}, $J^\text{Braun}_u$" and now trains on {\it real} data $\mathcal{D}^\text{real}_\text{DNS}$, which comprises the real recordings from the preliminary test datasets of DNS3 and both preliminary and blind test datasets of the Interspeech 2020 DNS Challenge (DNS1) \cite{reddy2020interspeechfinal} and the ICASSP 2020 DNS Challenge (DNS2) \cite{reddy2021icassp}. For both EP- and MB-based alternating training protocols, the {\tt FCRN} DNS is trained on $\mathcal{D}^\text{real}_\text{DNS}$ for $26$ epochs, while the {\tt PESQ-DNN} is updated employing the {\it synthetic} dataset $\mathcal{D}^\text{train}_\text{DNS3}$, with learning rates of $2\cdot10^{-5}$ and $5\cdot10^{-5}$, respectively. Our newly trained {\tt FCRN} DNS with different training protocols are marked with either ``EP" or ``MB" in the following result tables. As another reference, we run the \nth{1}-stage fine-tuning for additional $26$ epochs based on ``{\tt FCRN}, $J^\text{Braun}_u$", thus illustrating that the performance improvement in the \nth{2}-stage fine-tuning cannot solely be attributed to more training epochs. This reference is denoted as ``{\tt FCRN}, $J^\text{Braun}_u$, cont." in Tabs.\,\ref{DNS1_dev}, \ref{DNS2_test}, and \ref{DNS3_dev}.

As further baselines employed in Tabs.\,\ref{DNS1_dev}, \ref{DNS2_test}, and \ref{DNS3_dev}, we follow \cite{xu2021inter,xu2021deepT,xu2022iwaenc} and perform the \nth{2}-stage fine-tuning on ``{\tt FCRN} \cite{strake2020DNS}" with either the non-intrusive {\tt PESQNet} or the intrusive {\tt MF-PESQNet}. The baseline ``{\tt FCRN}/{\tt PESQNet} \cite{xu2021inter}" represents the {\tt FCRN} DNS fine-tuned on both {\it synthetic} dataset $\mathcal{D}^\text{train}_\text{DNS3}$ and {\it real} dataset $\mathcal{D}^\text{real}_\text{DNS}$ employing {\tt PESQNet}, but with a joint training protocol proposed in \cite{xu2021inter}, revealed to be unstable. Baselines ``{\tt FCRN}/{\tt PESQNet} \cite{xu2021deepT}" and ``{\tt FCRN}/{\tt MF-PESQNet} \cite{xu2022iwaenc}" represent the {\tt FCRN} model from \cite{xu2021deepT, xu2022iwaenc}, which are fine-tuned employing the old {\tt PESQNet} and {\tt MF-PESQNet}, respectively, using the (successful) EP-based alternating training protocols on {\it synthetic} dataset $\mathcal{D}^\text{train}_\text{DNS3}$. Furthermore, we include the DNS3 baseline \cite{braun2020data} provided by the challenge organizers, named as ``DNS3 Baseline \cite{braun2020data}".

We use the preliminary {\it synthetic} test set $\mathcal{D}^\text{dev}_\text{DNS1}$ from DNS1 \cite{reddy2020interspeechfinal} for development, see results in Tab.\,\ref{DNS1_dev}. The final evaluation is reported on both the {\it synthetic} test set $\mathcal{D}^\mathrm{test}_\mathrm{DNS2}$ from DNS2 (Tab.\,\ref{DNS2_test}) and the {\it real} blind test set $\mathcal{D}^\mathrm{test}_\mathrm{DNS3}$ from DNS3 (Tab.\,\ref{DNS3_dev}). We employ intrusive instrumental metrics to evaluate the performance on {\it synthetic} test data, including PESQ \cite{ITUT_pesq_wb_corri}, STOI \cite{taal2010short}, segmental SNR improvement $\Delta\text{SNR}_\text{seg}$ \cite{loizou2013speech}, and speech-to-reverberation modulation energy ratio (SRMR) \cite{falk2010non}. Please note that $\Delta\text{SNR}_\text{seg}$ is only employed under the conditions without reverberations to explicitly evaluate the denoising performance, while SRMR reflects the dereverberation effects and is only measured under reverberated conditions. We also employ the non-intrusive instrumental measure of the latest DNSMOS \cite{reddy2022dnsmos} on both {\it synthetic} and {\it real} test data, reporting either the overall (OVRL) enhanced speech quality or the detailed ones, including speech component quality (SIG) and background noise (BAK).
\vspace*{-2mm}
\subsection{Results and Discussion}
\vspace*{-1mm}
\begin{table}[t!]
\centering
\caption{Results on the \textbf{synthetic test set} $\mathcal{D}^\mathrm{test}_\mathrm{DNS2}$. DNSMOS refer to the overall (OVRL) metric. Best results are in {\bf bold} font, and the second best are \underline{underlined}. Proposed method with $*$}
\ninept
\setlength\tabcolsep{2pt}
\vspace*{-2mm}
\begin{adjustbox}{width=0.99\columnwidth,center}
	\begin{tabular}{c c c c c c c}
		\hline
		& \multirow{2}{*}{Methods} & \multicolumn{2}{c}{\stz{DNS fine-tuning}} & \multirow{2}{*}{PESQ} & \multirow{2}{*}{DNSMOS}  &  \multirow{2}{*}{STOI}\\ \cmidrule(l{.25em}r{.25em}){3-4}
		&   & \multicolumn{1}{c}{syn} & \multicolumn{1}{c}{real} &  &  &   \\ \hline
		\multirow{2}{*}{}& \multicolumn{1}{l}{\stz{Clean}} &    -  &  -  & \stz{4.64} & 3.28 & 1.00\\ \hhline{~~~~~~~}
		& \multicolumn{1}{l}{\stz{Noisy}} &    -  &  -  & \stz{2.37} & 2.54 & 0.88\\ \hline
		\multirow{8}{*}{\rotatebox{90}{REF \quad}}& \multicolumn{1}{l}{\stz{DNS3 Baseline \cite{braun2020data}}}&    \cmark  &  \xmark & \stz{3.14} & 2.99 & \underline{0.91}\\ \hhline{~~~~~~~}
		& \multicolumn{1}{l}{\stz{{\tt FCRN} \cite{strake2020DNS}}}&    \cmark  &  \xmark & 3.25 & 3.00 & \underline{0.93}\\ \hhline{~~~~~~~}
		& \multicolumn{1}{l}{\stz{{\tt FCRN}/{\tt PESQNet} \cite{xu2021inter}}}&    \cmark  &  \cmark & \stz{3.24} & {3.06} & \underline{0.93}\\ \hhline{~~~~~~~}
		& \multicolumn{1}{l}{\stz{{\tt FCRN}/{\tt PESQNet}, EP \cite{xu2021deepT}}} &    \cmark  &  \xmark& \stz{3.34} & 3.02 & \underline{0.93}\\ \hhline{~~~~~~~}
		& \multicolumn{1}{l}{\stz{{\tt FCRN}/{\tt MF-PESQNet}, EP \cite{xu2022iwaenc}}}&    \cmark  &  \xmark & \stz{3.37} & 3.03& \underline{0.93}\\ \hhline{~~~~~~~}
		& \multicolumn{1}{l}{\stz{{\tt FCRN}, $J^\text{Braun}$}}&    \cmark  &  \xmark & \underline{3.42} & {3.06} & {\bf 0.94}\\ \hhline{~~~~~~~}
		& \multicolumn{1}{l}{\stz{{\tt FCRN}, $J^\text{Braun}$, cont.}}&    \cmark  &  \xmark & 3.39 & 3.05 & {\bf 0.94}\\ \hline
		\multirow{2}{*}{\rotatebox{90}{NEW}}& \multicolumn{1}{l}{\stz{{\tt FCRN}/{\tt PESQ-DNN}, MB}}&    \xmark  &  \cmark & 3.18 & \underline{3.07} & \underline{0.93}\\ \hhline{~~~~~~~}
		& \multicolumn{1}{l}{\stz{{\tt FCRN}/{\tt PESQ-DNN}, EP$^*$}}&    \xmark  &  \cmark  & \stz{\bf 3.46} & {\bf 3.11} & {\bf 0.94}\\ \hline
	\end{tabular}
\end{adjustbox}
\label{DNS2_test}
\vspace*{-3mm}
\end{table}
In Tab.\,\ref{DNS1_dev}, we evaluate all methods on the {\it synthetic} development set $\mathcal{D}^\mathrm{dev}_\mathrm{DNS1}$ and report them separately under the conditions with and without reverberation. We employ intrusive instrumental metrics and the OVRL metric of DNSMOS. The types of data used for fine-tuning (or training) the DNS model may be synthetic (``syn") and/or real (``real"), as given in Tabs.\,\ref{DNS1_dev}, \ref{DNS2_test}, and \ref{DNS3_dev}. Our proposed method is marked with $*$, selected by providing the best development performance from Tab.\,\ref{DNS1_dev}. Our first observation is that the {\tt FCRN} DNS topology is a good choice, since it exceeds the DNS3 Baseline in all listed metrics. Comparing all {\tt FCRN} reference (REF) methods that are purely trained with synthetic data, we find the intrusive {\tt PESQNet} approach ``{\tt FCRN}/{\tt MF-PESQNet}, EP \cite{xu2022iwaenc}" to be a strong method, but the simple {\tt FCRN} trained with the loss \eqref{Braun} by Braun et al.\ has most first and second ranks. Analyzing our two new proposed schemes (last two rows, real data has been used in an extra fine-tuning step), we find out that the alternation on epoch basis (EP) performs roughly the same as ``{\tt FCRN}, $J^\text{Braun}$" in all metrics, and is much better than the minibatch (MB) alternation. To be fair, we also spent a longer fine-tuning for ``{\tt FCRN}, $J^\text{Braun}$" (marked with ``cont."), but the reference method started to degrade. Please note that our proposed EP method excels the only earlier approach using real data as well ({\tt FCRN}/{\tt PESQNet} \cite{xu2021inter}) by a large margin (see, e.g., $3.60$ vs. $3.29$ PESQ points). 

In Tab.\,\ref{DNS2_test}, we report the performance of all the investigated methods measured on the {\it synthetic} test set $\mathcal{D}^\mathrm{test}_\mathrm{DNS2}$, employing PESQ, STOI, and the OVRL DNSMOS metric. On the test data, we see among the reference methods about the same rank orders as on the development set in Tab.\,\ref{DNS1_dev}. However, our proposed ``{\tt FCRN}/{\tt PESQ-DNN}, EP$^*$" method excels the strong approach ``{\tt FCRN}/{\tt MF-PESQNet}, EP \cite{xu2022iwaenc}" and ``{\tt FCRN}, $J^\text{Braun}$" with $3.46$ vs.\ $3.37$/$3.42$ (PESQ) and $3.11$ vs.\ $3.06$/$3.03$ (DNSMOS), which is an $0.05$ points DNSMOS improvement. {\it Note that the improvement vs.\ the DNS3 Baseline is even $0.12$ DNSMOS points. Our ``{\tt FCRN}/{\tt PESQ-DNN}, EP$^*$" method exceeds the DNS3 Baseline by a highly significant $0.32$ PESQ points.}

Tab.\,\ref{DNS3_dev} presents the performance of all the investigated methods on the {\it real} test set $\mathcal{D}^\mathrm{test}_\mathrm{DNS3}$. Due to the lack of a clean reference signal, we only measure the non-intrusive DNSMOS and separately report the SIG, BAK, and OVRL dimension scores. On real test data, we observe that there are indeed outstanding reference methods for background noise suppression and residual noise quality (BAK), as both ``{\tt FCRN}/{\tt PESQNet} \cite{xu2021inter}" (partly using real data for fine-tuning) and ``{\tt FCRN}, $J^\text{Braun}$" reach a score of $3.90$ in that quality dimension. They are also best among the reference methods in the OVRL DNSMOS ($2.61$ points). {\it Note that our proposed ``{\tt FCRN}/{\tt PESQ-DNN}, EP$^*$" approach is best in all DNSMOS scores, reaching an $0.05$ OVRL DNSMOS score higher, as it employs real data for fine-tuning in an advantageous manner.}

\begin{table}[t!]
\centering
\caption{DNSMOS \cite{reddy2022dnsmos} results on the \textbf{real test set} $\mathcal{D}^\mathrm{test}_\mathrm{DNS3}$. Best results are in {\bf bold} font, and the second best are \underline{underlined}. Proposed method with $*$}
\setlength\tabcolsep{3pt}
\vspace*{-2mm}
\begin{adjustbox}{width=0.956\columnwidth,center}
	\begin{tabular}{c c c c c c c}
		\hline
		&\multirow{2}{*}{Methods} & \multicolumn{2}{c}{\stz{DNS fine-tuning}}& \multicolumn{3}{c}{\stz{DNSMOS}}\\ \cmidrule(l{.25em}r{.25em}){3-4} \cmidrule(l{.25em}r{.25em}){5-7}
		&   & \multicolumn{1}{c}{syn} & \multicolumn{1}{c}{real}&  SIG & BAK & OVRL\\ \hline
		& \multicolumn{1}{l}{\stz{Noisy}}&    -  &  - & 2.89 & 2.34 & 2.11\\ \hline
		\multirow{7}{*}{\rotatebox{90}{REF \quad}}& \multicolumn{1}{l}{\stz{DNS3 Baseline \cite{braun2020data}}}&    \cmark  &  \xmark & \underline{2.90} & 3.80 & 2.60\\ \hhline{~~~~~~~}
		& \multicolumn{1}{l}{\stz{{\tt FCRN} \cite{strake2020DNS}}}&    \cmark  &  \xmark & 2.86 & 3.71 & 2.55\\ \hhline{~~~~~~~}
		& \multicolumn{1}{l}{\stz{{\tt FCRN}/{\tt PESQNet} \cite{xu2021inter}}}&    \cmark  &  \cmark & 2.87 & \underline{3.90} & \underline{2.61}\\ \hhline{~~~~~~~}
		& \multicolumn{1}{l}{\stz{{\tt FCRN}/{\tt PESQNet}, EP \cite{xu2021deepT}}}&    \cmark  &  \xmark & 2.85 & 3.78 & 2.56\\ \hhline{~~~~~~~}
		& \multicolumn{1}{l}{\stz{{\tt FCRN}/{\tt MF-PESQNet}, EP \cite{xu2022iwaenc}}}&    \cmark  &  \xmark & 2.86 & 3.79 & 2.57\\ \hhline{~~~~~~~}
		& \multicolumn{1}{l}{\stz{{\tt FCRN}, $J^\text{Braun}$}}&    \cmark  &  \xmark & 2.86 & \underline{3.90} & \underline{2.61}\\ \hhline{~~~~~~~}
		& \multicolumn{1}{l}{\stz{{\tt FCRN}, $J^\text{Braun}$, cont.}}&    \cmark  &  \xmark & 2.87 & 3.88 & \underline{2.61}\\ \hline
		\multirow{2}{*}{\rotatebox{90}{NEW}}& \multicolumn{1}{l}{\stz{{\tt FCRN}/{\tt PESQ-DNN}, MB}}&    \xmark  &  \cmark & 2.88 & 3.82 & 2.60\\ \hhline{~~~~~~~}
		& \multicolumn{1}{l}{\stz{{\tt FCRN}/{\tt PESQ-DNN}, EP$^*$}}&    \xmark  &  \cmark  & {\bf 2.91} & {\bf 3.95} & {\bf 2.66}\\ \hline		
	\end{tabular}
\end{adjustbox}
\label{DNS3_dev}
\vspace*{-3mm}
\end{table}
\section{Conclusions}
\vspace*{-1mm}
In this work, we show how to fine-tune a deep noise suppression (DNS) model with {\it real} data, with the help of an end-to-end non-intrusive deep neural network (DNN) named {\tt PESQ-DNN}, which estimates perceptual evaluation of speech quality (PESQ) scores of the enhanced speech signal. The employed {\tt PESQ-DNN} provides a differentiable reference-free perceptual loss for the DNS training, aiming at maximizing the PESQ scores of the trained DNS model. An epoch-level alternating training protocol to train the DNS model with one epoch of {\it real} data, followed by {\tt PESQ-DNN} updating with one epoch of {\it synthetic} data, turned out to be strongest. Detailed analyses show that the DNS model trained with the {\tt PESQ-DNN} and real data outperforms all reference methods employing only synthetic data. On synthetic test data, our proposed method excels the Interspeech 2021 DNS Challenge baseline by a highly significant $0.32$ PESQ points. Both on synthetic and on real test data, our baseline (not employing real training data) is exceeded by the proposed method (employing real training data) by $0.05$ DNSMOS points -- although the employed \texttt{PESQ-DNN} optimizes for PESQ.

\bibliographystyle{IEEEbib}%
\bibliography{mainTrans21}

\begin{thebibliography}{10}

\bibitem{pascual2017segan}
{S}. {P}ascual, {A}. {B}onafonte, and {J}. {S}erra,
\newblock ``{SEGAN: Speech Enhancement Generative Adversarial Network},''
\newblock {\em arXiv preprint arXiv:1703.09452}, Jun. 2017.

\bibitem{pandey2018adversarial}
{A}. {P}andey and {D}.~{L}. {W}ang,
\newblock ``{On Adversarial Training and Loss Functions for Speech
  Enhancement},''
\newblock in {\em {P}roc. of {ICASSP}}, Calgary, AB, Canada, Apr. 2018, pp.
  5414--5418.

\bibitem{wang2018supervised}
{D}.~{L}. {W}ang and {J}.~{T}. {C}hen,
\newblock ``{S}upervised {S}peech {S}eparation {B}ased on {D}eep {L}earning:
  {A}n {O}verview,''
\newblock {\em IEEE/ACM T-ASLP}, vol. 26, no. 10, pp. 1702--1726, Oct. 2018.

\bibitem{tan2019complex}
{K}. {T}an and {D}.~{L}. {W}ang,
\newblock ``{C}omplex {S}pectral {M}apping with a {C}onvolutional {R}ecurrent
  {N}etwork for {M}onaural {S}peech {E}nhancement,''
\newblock in {\em {P}roc. of {ICASSP}}, Brighton, UK, May 2019, pp. 6865--6869.

\bibitem{strake2019separated}
{M}. {S}trake, {B}. {D}efraene, {K}. {F}luyt, {W}. {T}irry, and {T}.
  {Fing\-scheidt},
\newblock ``{S}eparated {N}oise {S}uppression and {S}peech {R}estoration:
  {LSTM}-{B}ased {S}peech {E}nhancement in {T}wo {S}tages,''
\newblock in {\em {P}roc. of {WASPAA}}, New Paltz, NY, USA, Oct. 2019, pp.
  239--243.

\bibitem{strake2020fully}
{M}. {S}trake, {B}. {D}efraene, {K}. {F}luyt, {W}. {T}irry, and {T}.
  {Fing\-scheidt},
\newblock ``{F}ully {C}onvolutional {R}ecurrent {N}etworks for {S}peech
  {E}nhancement,''
\newblock in {\em {P}roc. of {ICASSP}}, Barcelona, Spain, May 2020, pp.
  6674--6678.

\bibitem{strake2020DNS}
{M}. {S}trake, {B}. {D}efraene, {K}. {F}luyt, {W}. {T}irry, and {T}.
  {Fing\-scheidt},
\newblock ``{INTERSPEECH} 2020 {D}eep {N}oise {S}upression {C}hallenge: {A}
  {F}ully {C}onvolutional {R}ecurrent {N}etwork {(FCRN)} for {J}oint
  {D}ereverberation and {D}enoising,''
\newblock in {\em {P}roc. of {INTERSPEECH}}, Shanghai, China, Oct. 2020, pp.
  2467--2471.

\bibitem{braun2021consolidated}
{S}. {B}raun and {I}. {T}ashev,
\newblock ``{A Consolidated View of Loss Functions for Supervised Deep
  Learning-Based Speech Enhancement},''
\newblock in {\em {P}roc. of {TSP}}, Brno, Czech Republic, Jul. 2021, pp.
  72--76.

\bibitem{ITUT_pesq_wb_corri}
{I}{T}{U},
\newblock {\em {Rec. P.862.2: Corrigendum 1, Wideband Extension to
  Recommendation P.862 for the Assessment of Wideband Telephone Networks and
  Speech Codecs}},
\newblock {International Telecommunication Standardization Sector (ITU-T)},
  Oct. 2017.

\bibitem{ITUT_polqa_2018}
{I}{T}{U},
\newblock {\em {Rec. P.863: Perceptual Objective Listening Quality Prediction
  (POLQA)}},
\newblock {International Telecommunication Union, Telecommunication
  Standardization Sector (ITU-T)}, Mar. 2018.

\bibitem{taal2010short}
{C}.~{H}. Taal, {R}.~{C}. {H}endriks, {R}. {H}eusdens, and {J}. {J}ensen,
\newblock ``{A} {S}hort-{T}ime {O}bjective {I}ntelligibility {M}easure for
  {T}ime-{F}requency {W}eighted {N}oisy {S}peech,''
\newblock in {\em {P}roc. of {ICASSP}}, Dallas, TX, USA, Jun. 2010, pp.
  4214--4217.

\bibitem{reddy2021icassp}
{C}. {K}.~{A}. {R}eddy, {H}. {D}ubey, {V}. {G}opal, {R}. {C}utler, {S}.
  {B}raun, {H}. {G}amper, {R}. {A}ichner, and {S}. {S}rinivasan,
\newblock ``{ICASSP 2021 Deep Noise Suppression Challenge},''
\newblock in {\em {P}roc. of {ICASSP}}, Toronto, ON, Canada, Jun. 2021, pp.
  6623--6627.

\bibitem{reddy2021interspeechIN}
{C}. {K}.~{A}. {R}eddy, {H}. {D}ubey, {K}. {K}oishida, {A}. {N}air, {V}.
  {G}opal, {R}. {C}utler, {S}. {B}raun, {H}. {G}amper, {R}. {A}ichner, and {S}.
  {S}rinivasan,
\newblock ``{INTERSPEECH 2021 Deep Noise Suppression Challenge},''
\newblock in {\em {P}roc. of {INTERSPEECH}}, Brno, Czech Republic, Aug. 2021,
  pp. 2796--2800.

\bibitem{reddy2021dnsmos}
{C}. {K}.~{A}. {R}eddy, {V}. {G}opal, and {R}. {C}utler,
\newblock ``{DNSMOS: A Non-Intrusive Perceptual Objective Speech Quality Metric
  to Evaluate Noise Suppressors},''
\newblock in {\em {P}roc. of {ICASSP}}, Toronto, ON, Canada, Jun. 2021, pp.
  6493--6497.

\bibitem{ITU_P808}
{I}{T}{U},
\newblock {\em {Rec. P.808: Subjevtive Evaluation of Speech Quality With a
  Crowdsoucing Approach}},
\newblock {International Telecommunication Standardization Sector (ITU-T)},
  Feb. 2018.

\bibitem{reddy2022dnsmos}
{C}. {K}.~{A}. {R}eddy, {V}. {G}opal, and {R}. {C}utler,
\newblock ``{DNSMOS P. 835: A Non-Intrusive Perceptual Objective Speech Quality
  Metric to Evaluate Noise Suppressors},''
\newblock in {\em {P}roc. of {ICASSP}}, Singapore, Singapore, Apr. 2022, pp.
  886--890.

\bibitem{ITUT_P835}
{I}{T}{U},
\newblock {\em {Rec. P.835: Corrigendum 1, Subjective Test Methodology for
  Evaluating Speech Communication Systems that Include Noise Suppression
  Algorithm}},
\newblock {International Telecommunication Standardization Sector (ITU-T)},
  Jan. 2011.

\bibitem{mittag2021nisqa}
{G}. {M}ittag, {B}. {N}aderi, {A}. {C}hehadi, and {S}. {M}{\"o}ller,
\newblock ``{NISQA: A Deep CNN-Self-Attention Model for Multidimensional Speech
  Quality Prediction with Crowdsourced Datasets},''
\newblock in {\em {P}roc. of {INTERSPEECH}}, Brno, Czech Republic, Aug. 2021,
  pp. 2127--2131.

\bibitem{xu2021inter}
{Z}. {X}u, {M}. {S}trake, and {T}. {F}ingscheidt,
\newblock ``{Deep Noise Suppression With Non-Intrusive PESQNet Supervision
  Enabling the Use of Real Training Data},''
\newblock in {\em {P}roc. of {INTERSPEECH}}, Brno, Czech Republic, Aug. 2021,
  pp. 2806--2810.

\bibitem{xu2021deepT}
{Z}. {X}u, {M}. {S}trake, and {T}. {Fing\-scheidt},
\newblock ``{Deep Noise Suppression Maximizing Non-Differentiable PESQ Mediated
  by a Non-Intrusive PESQNet},''
\newblock {\em IEEE/ACM T-ASLP}, vol. 30, no. 4, pp. 1572--1585, Apr. 2022.

\bibitem{xu2022iwaenc}
{Z}. {X}u, {M}. {S}trake, and {T}. {F}ingscheidt,
\newblock ``{Does a PESQNet (Loss) Require a Clean Reference Input? The
  Original PESQ Does, But ACR Listening Tests Don't},''
\newblock in {\em {P}roc. of {IWAENC}}, Bamberg, Germany, Sep. 2022, pp. 1--5.

\bibitem{xu2023PESQDNN}
{Z}. {X}u, {Z}. {Z}hao, and {T}. {F}ingscheidt,
\newblock ``{Coded Speech Quality Measurement by a Non-Intrusive PESQ-DNN},''
\newblock {\em arXiv preprint arXiv: 2304.09226}, Apr. 2023.

\bibitem{braun2020data}
{S}. {B}raun and {I}. {T}ashev,
\newblock ``{Data Augmentation and Loss Normalization for Deep Noise
  Suppression},''
\newblock in {\em {P}roc. of {SPECOM}}, St. Petersburg, Russia, Oct. 2020, pp.
  79--86.

\bibitem{Garofalo2007}
{J}. {G}arofolo, {D}. {G}raff, {D}. {P}aul, and {D}. {P}allett,
\newblock ``{CSR-I} ({WSJ0}) {C}omplete,''
\newblock {\em {Linguistic Data Consortium, Philadelphia}}, 2007.

\bibitem{thiemann2013diverse}
{J}. {T}hiemann, {N}. {I}to, and {E}. {V}incent,
\newblock ``{The Diverse Environments Multi-Channel Acoustic Noise Database: A
  Database of Multichannel Environmental Noise Recordings},''
\newblock {\em {J. Acoustic. Soc. Am.}}, vol. 133, no. 5, pp. 3591--3591, 2013.

\bibitem{dean2010qut}
{D}.~{B}. {D}avid, {S}. {S}ridharan, {R}.~{J}. {V}ogt, and {M}.~{W}. {M}ason,
\newblock ``{The QUT-NOISE-TIMIT Corpus for the Evaluation of Voice Activity
  Detection Algorithms},''
\newblock in {\em {P}roc. of {INTERSPEECH}}, Makuhari, Japan, Sept. 2010, pp.
  3110--3113.

\bibitem{reddy2020interspeechfinal}
{C}. {K}.~{A}. {R}eddy, {H}. {D}ubey, {V}. {G}opal, {R}. {C}heng, {R}.
  {C}utler, {S}. {M}atusevych, {R}. {A}ichner, {A}. {A}azami, {S}. {B}raun,
  {P}. {R}ana, {S}. {S}rinivasan, and {J}. {G}ehrke,
\newblock ``{The Interspeech 2020 Deep Noise Suppression Challenge: Datasets,
  Subjective Testing Framework, and Challenge Results},''
\newblock in {\em {P}roc. of {INTERSPEECH}}, Shanghai, China, Oct. 2020, pp.
  2492--2496.

\bibitem{loizou2013speech}
{P}.~{C}. {L}oizou,
\newblock {\em {Speech Enhancement: Theory and Practice}},
\newblock CRC press, 2013.

\bibitem{falk2010non}
{T}.~{H}. {F}alk, {C}. {Z}heng, and {W}.~{Y}. Chan,
\newblock ``{A Non-Intrusive Quality and Intelligibility Measure of Reverberant
  and Dereverberated Speech},''
\newblock {\em IEEE T-ASLP}, vol. 18, no. 7, pp. 1766--1774, Aug. 2010.

\end{thebibliography}

\end{document}